\documentclass{PoS}
\usepackage{amsmath,amsfonts, amssymb,bbm}
\title{Breakdown of QCD coherence ?}

\ShortTitle{Breakdown of coherence?}

\author{\speaker{A. Kyrieleis}\\
        University of Manchester, U.K.\\
        E-mail: \email{kyrieleis@hep.man.ac.uk}}

\author{J.R. Forshaw\\
        University of Manchester, U.K.\\
        E-mail: \email{forshaw@mail.cern.ch}}
\author{M.H. Seymour\\
        CERN and University of Manchester, U.K.\\
        E-mail: \email{mike.seymour@cern.ch}}

\abstract{We reconsider the calculation of a non-global QCD observable and find the possible breakdown of QCD coherence. This breakdown arises as a result of wide angle soft gluon emission developing a sensitivity to emission at small angles and it leads to the appearance of super-leading logarithms. We use the `gaps between jets' cross-section as a concrete example and illustrate that the new logarithms are intimately connected with the presence of Coulomb gluon contributions.  Numerical estimates of their potential phenomenological significance are presented.
}

\FullConference{International Workshop on Diffraction in High-Energy Physics -DIFFRACTION 2006 -\\
                 September 5-10 2006\\
                 Adamantas, Milos island, Greece}

\def\Journal#1#2#3#4{{#1} {\bf #2}, #3 (#4)}

\def\NPB{{\em Nucl. Phys.} B}
\def\PLB{{\em Phys. Lett.} B}
\def\PRL{\em Phys. Rev. Lett.}
\def\PRD{{\em Phys. Rev.} D}

\def\JHP{\em JHEP}
\def\EJC{{\em Eur. Phys. J.} C}

\def\be{\begin{equation}}
\def\ee{\end{equation}}
\def\bea{\begin{eqnarray}}
\def\eea{\end{eqnarray}}

\begin{document}
\section{Introduction}
The resummation of large logarithms associated with wide angle soft gluon
emissions has been investigated for the last 20 years. 
For certain observables the  contributions from non-global
logarithms \cite{NGL}  have to be taken into account.
%The 
%discovery of non-global logarithms \cite{NGL} established the class
%of non-observables - which are sensitive to these logarithms.
%of those observables that receive a non-suppressed contribution from these
%logarithms - the non-global observables.
One of the simplest of these non-global observables is the
`gaps-between-jets' cross-section. This is the cross-section for
producing a pair of high transverse momentum jets (Q) with a restriction
on the transverse momentum of any additional jets radiated in between
the two jets, i.e. $k_T < Q_0$ for emissions in the gap. This
observable has been studied \cite{OdSter,AppSey} and has been measured
at HERA and the Tevatron \cite{gaps exp}. 

\begin{figure}[h]
\begin{center}
\includegraphics[height=2in]{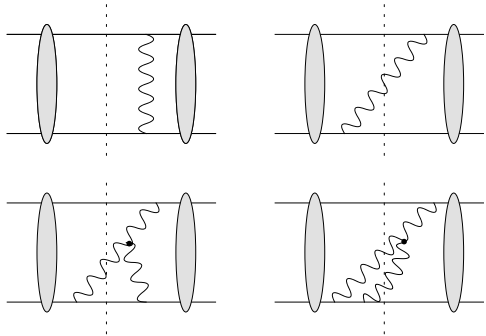}\\[0pt]
\end{center}
\caption{Illustrating the cancellation (and miscancellation) of soft gluon
corrections.}%
\label{fig:miscancel}%
\end{figure}

In the original calculations \cite{OdSter} of the gaps-between-jets
cross section,  all terms proportional to
$\alpha_{s}^{n}\ln^{n}(Q^2/Q_{0}^2)$  that can be obtained by dressing the primary $2\rightarrow2$ scattering in all possible ways with soft virtual gluons were summed.  The restriction to soft gluons implies the use of the
eikonal approximation. Let us  focus on   quark-quark scattering from
now on. The corresponding resummed cross-section can be written
\begin{equation}
\sigma=\mathbf{M}^{\dag}\mathbf{S}_{V}\mathbf{M}\qquad\mbox{with}\qquad
\mathbf{M}=\exp\left(  -\frac{2\alpha_{s}}{\pi}%
%TCIMACRO{\dint \limits_{Q_{0}}^{Q}}%
%BeginExpansion
{\displaystyle\int\limits_{Q_{0}}^{Q}}
%EndExpansion
\frac{dk_{T}}{k_{T}}\Gamma\right)  \mathbf{M}_0.\label{eq:OS}%
\end{equation}
Here, $\mathbf{M}$ is the all-orders $qq\rightarrow qq$  amplitude (a 2-component vector in colour space), $\mathbf{M}_0$ is the hard scattering
amplitude and $\mathbf{S}_V$ represents the cut.
% and $S_V$ is a 2x2 matrix. 
The anomalous dimension matrix
$\Gamma$ \cite{Gamma} incorporates the effect of dressing a
$qq\rightarrow qq$
amplitude with a  virtual gluon in all possible ways.
It receives contributions from  two distinct regions of the
loop-integral: the first corresponds to an on-shell gluon (to which
one can  assign a rapidity) and is
identical, but with  opposite sign, to the contribution from a
real gluon.
The second contribution, sometimes referred to as the `Coulomb gluon
 contribution' \cite{Coulomb}  is
purely imaginary ($i\pi$ terms)  and stems from the region where the emitting parton is
on-shell.
 Eq. (\ref{eq:OS}) therefore corresponds to the  independent
 emission of soft gluons, i.e. the iterative dressing of the
 $2\rightarrow 2$ process with a softer gluon:  due to perfect real/virtual
 cancellation  outside the gap (the first line of Fig. \ref{fig:miscancel}
 shows two contributions) one only has to consider virtual gluons
 in the gap and the Coulomb terms.
% which are incorporated in $\Gamma$.

However, there is another source of leading logarithms. Let us
consider the two diagrams in the second line of
Fig.  \ref{fig:miscancel}. A real gluon (which is outside the gap by the
definition of our observable) emits a softer real or virtual
gluon. The real-virtual cancellation is guaranteed only for the softest
gluon. Since real gluons above $Q_0$ are forbidden in the gap, the
two diagrams do not completely cancel; the left diagram with the virtual gluon
being in the gap and its $k_T$ being larger than $Q_0$ survives. The
non-global nature of our observable has prevented the soft gluon
cancellation which is necessary in order that Eq. (\ref{eq:OS}) should be
the complete result. 

It is therefore necessary to include the emission of any
number of soft gluons outside the gap region (real and virtual)
dressed with any number of virtual gluons within the gap
region. Clearly it is a formidable challenge to sum all leading
logarithms, mainly because of the complicated colour
structure. Progress has been made, working in the large $N$
approximation \cite{AppSey}. Here, we keep the exact colour structure
but instead we only compute the cross-section for one gluon outside
the gap region. This can be viewed as the first term in an expansion
in the number of out-of-gap-gluons.
\section{Super-leading logarithms}
In order to extract the leading logarithms we consider soft gluons
strongly ordered in transverse momentum.
The cross-section for one gluon outside and any number
of gluons inside the gap  is split into two parts corresponding to a
virtual or real  out-of-gap gluon: 
\begin{align}
&\sigma_1    =-\bar\alpha\int_{Q_{0}}^{Q}\frac{dk_{T}}{k_{T}%
}~\int\limits_{\text{out}}\frac{dy~d\phi}{2\pi}\:\left(\Omega_V+
  \Omega_R\right)\:,\qquad\bar\alpha\equiv \frac{2\alpha_s}{\pi}
\label{eq:sig1}
\end{align}
%The two parts read:
\begin{align}
 \Omega_R= \mathbf{M}_{0}^{\dag}\exp\left(  { -}\bar\alpha%
\int\limits_{k_{T}}^{Q}\frac{dk_{T}^{\prime}}{k_{T}^{\prime}}\Gamma
^{\dag}\right)  &\mathbf{D}_{\mu}^{\dag}\exp\left(  { -}\bar\alpha\int\limits_{Q_{0}}^{k_{T}}\frac{dk_{T}^{\prime}}%
{k_{T}^{\prime}}\Lambda^{\dag}\right)  \mathbf{S}_{R}
\nonumber\\
 \exp\left(  { -}\bar\alpha\int\limits_{Q_{0}}^{k_{T}%
}\frac{dk_{T}^{\prime}}{k_{T}^{\prime}}\Lambda\right)  \mathbf{D}&
^{\mu}\exp\left(  { -}\bar\alpha\int\limits_{k_{T}}%
^{Q}\frac{dk_{T}^{\prime}}{k_{T}^{\prime}}\Gamma\right)
\mathbf{M}_{0}\;,\label{eq:real}\\
%\end{align}
%and for a virtual out-of-gap gluon%
%\begin{align}
%\sigma_{V}  &  =-\frac{2\alpha_{s}}{\pi}\int_{Q_{0}}^{Q}\frac{dk_{T}}{k_{T}%
%}\int\limits_{\text{out}}\frac{dy~d\phi}{2\pi}\quad\\
\Omega_V=   \mathbf{M}_{0}^{\dag}\exp\left({ -}\bar \alpha \int\limits_{Q_{0}}^{Q}\frac{dk_{T}^{\prime}}{k_{T}^{\prime}}\Gamma
^{\dag}\right)  \mathbf{S}_{V} \exp&\left({ -}\bar \alpha
\int\limits_{Q_{0}}^{k_{T}}\frac{dk_{T}^{\prime}}{k_{T}^{\prime}}\Gamma\right)
~\gamma\mathbf{~}\exp\left({ -}\bar\alpha\int\limits_{k_{T}}^{Q}\frac{dk_{T}^{\prime}}{k_{T}^{\prime}}\Gamma
\right)  \mathbf{M}_{0}~{\text +~}\text{{\text c.c.}}
\label{eq:virtual}%
\end{align}
$\mathbf{D}^\mu$ and $\boldsymbol{\gamma}$ are the matrices that  
represent the emission of a real and a virtual gluon ($k_T, y, \phi$)
outside the gap, respectively.  The major new ingredient is the matrix
$\Lambda$ \cite{LAM} which  incorporates the dressing
of the $qq\to qqg$ process with a virtual gluon. The emission of the
out-of-gap gluon is sandwiched between two exponentials: this accounts
for all possible positions of the out-of-gap gluon within a chain of
any number of $k_T$-ordered gluons within the gap. 

The phase space of the out-of-gap gluon in Eq. (\ref{eq:sig1})  includes the
configurations where it is collinear to either of the external
quarks. One might suppose that the corresponding divergences cancel
among $\Omega_R$ and $\Omega_V$. This is true in case of the final state
collinear limit. However, in the limit of the out-of-gap gluon
becoming collinear to one of the initial state quarks,
which corres\-ponds to $|y|\to\infty, k_T>Q_0$, there is no cancellation:
\begin{equation}
[\Omega_V+\Omega_R]_{|y|\to\infty} \ne 0.
\label{eq:oneout}
\end{equation}
In particular, $(\Omega_V+\Omega_R)$ becomes independent of $y$ in that limit.
This has severe consequences. As the out-of-gap region stretches to
infinity in  rapidity, the integral Eq. (\ref{eq:sig1}) is divergent as it
stands. This divergence however indicates that one needs to go beyond
the soft approximation when considering the out-of-gap gluon. 
%As it stands we have a divergence arising from the integral over the rapidity
%of the out-of-gap gluon:%
%\begin{equation}
%~\int\limits_{\text{out}}\frac{dy~d\phi}{2\pi}~\omega_{1}\sim y_{\text{max}%
%}-\frac{Y}{2}.\label{eq:rap}%
%\end{equation}
%In the soft approximation the integral is divergent which is the signal that
%we need to go beyond the soft approximation when considering these emissions.
Strictly speaking we ought to work in the collinear (but not soft)
approximation which means that the integral over rapidity ought to be replaced
by%
\begin{equation}
\int d^{2}k_{T}\int\limits_{\text{out}}dy~~\left.  \frac{d\sigma}{dyd^{2}%
k_{T}}\right\vert _{\text{soft}}\rightarrow\int d^{2}k_{T}\left[
\int\limits^{y_{\text{max}}}dy~\left.  \frac{d\sigma}{dyd^{2}k_{T}}\right\vert
_{\text{soft}}+\int\limits_{y_{\text{max}}}^{\infty}dy~\left.  \frac{d\sigma
}{dyd^{2}k_{T}}\right\vert _{\text{collinear}}\right]  .\label{eq:softcol}%
\end{equation}
In this equation $y_{\text{max}}$ is a matching point between the regions in
which the soft and collinear approximations are used. If $y_{\text{max}}$ is
in the region in which both approximations are valid the dependence on it
should cancel in the sum of the two terms. Now we know that%
\begin{equation}
\int\limits_{y_{\text{max}}}^{\infty}dy~\left.  \frac{d\sigma}{dyd^{2}k_{T}%
}\right\vert _{\text{collinear}}=\int\limits_{y_{\text{max}}}^{\infty
}dy~\left(  \left.  \frac{d\sigma_{\text{R}}}{dyd^{2}k_{T}}\right\vert
_{\text{collinear}}+\left.  \frac{d\sigma_{\text{V}}}{dyd^{2}k_{T}}\right\vert
_{\text{collinear}}\right)
\end{equation}
where the contribution due to real gluon emission can be written as%
\begin{align}
\int\limits_{y_{\text{max}}}^{\infty}dy~\left.  \frac{d\sigma_{\text{R}}%
}{dyd^{2}k_{T}}\right\vert _{\text{collinear}}~ 
%&  =\int\limits_{0}^{1-\delta
%}dz\frac{1}{2}\left(  \frac{1+z^{2}}{1-z}\right)  \frac{q(x/z,\mu^{2}%
%){q(x,\mu^{2})}A_{\text{R}}\nonumber\\
&  =\int\limits_{0}^{1-\delta}dz\frac{1}{2}\left(  \frac{1+z^{2}}{1-z}\right)
\left(  \frac{q(x/z,\mu^{2})}{q(x,\mu^{2})}-1\right)  A_{\text{R}}%
+\int\limits_{0}^{1-\delta}dz\frac{1}{2}\frac{1+z^{2}}{1-z}A_{\text{R}%
}\label{eq:col}%
\end{align}
and the contribution due to virtual gluon emission is%
\begin{equation}
\int\limits_{y_{\text{max}}}^{\infty}dy~\left.  \frac{d\sigma_{\text{V}}%
}{dyd^{2}k_{T}}\right\vert _{\text{collinear}}~=\int\limits_{0}^{1-\delta
}dz\frac{1}{2}\left(  \frac{1+z^{2}}{1-z}\right)  A_{\text{V}}\text{.}%
\end{equation}
In Eq. (\ref{eq:col}), $q(x,\mu^{2})$ is the parton distribution function for a
quark in a hadron at scale $\mu^{2}$ and momentum fraction $x$. The factors
$A_{\text{R}}$ and $A_{\text{V}}$ contain the $z$ independent factors which
describe the soft gluon evolution.  Since we require $y>y_{\text{max}}$\footnote{The approximation arises
since we assume for simplicity that $\Delta y$ is large and $\delta$ is small.
This approximation does not affect the leading behaviour and can easily be
made exact if necessary.} the upper limit on the $z$ integral is
fixed: $
%\begin{equation}
\delta\approx k_T/Q\cdot \exp (  y_{\text{max}}-\Delta y/2 ).
%\end{equation}
$
We have already established that $A_{\text{R}}+A_{\text{V}}\neq0$ due to
Coulomb gluon contributions to the evolution. If it were the case that
$A_{\text{R}}+A_{\text{V}}=0$ then the virtual emission contribution would
cancel identically with the corresponding term in the real emission
contribution leaving behind a term regularised by the `plus prescription'
(since we can safely take $\delta\rightarrow0$ in the first term of
Eq. (\ref{eq:col})). This term could then be absorbed into the evolution of the
incoming quark parton distribution function by choosing the factorisation
scale to equal the jet scale $Q$.

The miscancellation therefore induces an additional contribution of the form%
\begin{align}
\int\limits_{0}^{1-\delta}dz\frac{1}{2}\left(  \frac{1+z^{2}}{1-z}\right)
(A_{\text{R}}+A_{\text{V}}) &  =\ln\left(  \frac{1}{\delta}\right)
(A_{\text{R}}+A_{\text{V}})+\text{subleading}\\
&  \approx\left(  -y_{\text{max}}+\frac{\Delta y}{2}+\ln\left(  \frac{Q}%
{k_{T}}\right)  \right)  (A_{\text{R}}+A_{\text{V}}).
\end{align}
Provided we stay within the soft-collinear region in which both the soft and
collinear approximations are valid, the $y_{\text{max}}$ dependence will
cancel with that coming from the soft contribution in Eq. (\ref{eq:softcol})
leaving only the logarithm. The leading effect of treating properly the
collinear region is therefore simply to introduce an effective upper limit  $\Delta y/2+\ln(Q/k_{T})$ to the integration over rapidity in Eq. (\ref{eq:softcol}).
% More precisely, we can
%therefore estimate the leading behaviour simply by setting $y_{\text{max}%
%}=\Delta y/2+\ln(Q/k_{t})$ in the soft integral, effectively including the
%entire soft-collinear region. We are left with
\begin{equation}
~\frac{2\alpha_{s}}{\pi}\int_{Q_{0}}^{Q}\frac{dk_{T}}{k_{T}}\int
\limits_{Y/2}^{\ln(Q/k_{T})+\Delta y/2}\frac{dy~d\phi}{2\pi}= ~\frac
{2\alpha_{s}}{\pi}\frac{1}{2}\ln^{2}(Q/Q_{0})+\text{subleading}.
\end{equation}
This is the super-leading logarithm: the failure of the `plus prescription'
has resulted in the generation of an extra collinear logarithm. The
implications for the gaps-between-jets cross-section are clear: collinear
logarithms can be summed into the parton density functions only up to scale
$Q_{0}$ and the logarithms in $Q/Q_{0}$ from further collinear evolution must
be handled separately. 
%Moreover, since we now have a source of double
%logarithms, the calculation of the single logarithmic series necessarily
%requires knowledge of the two-loop evolution matrices.

The miscancellation Eq. (\ref{eq:oneout}) and hence the super-leading
logarithm is intimately connected with
the Coulomb phase terms. If one artificially switches off the $i \pi$
terms in the evolution matrices, then there is full cancellation  in
Eq. (\ref{eq:oneout}). Moreover, the super-leading logarithm makes its
appearance at the lowest possible order in $\alpha_s$, i.e. at 
$O(\alpha_s^4)$ relative to the Born cross-section. This is due to the fact that at lower orders any $i\pi$ term is cancelled by a corresponding  term from the complex conjugate contribution.
%More explicitly, the 
%$O(\alpha_{s})$ and $O(\alpha_{s}^{2})$ corrections to the Born cross-section simply never involve
%more than one $i\pi$ term and hence any $i\pi$ terms must cancel since the
%cross-section is real. The first candidate order at which two factors of
%$i\pi$ can appear is therefore $O(\alpha_{s}^{3})$. However, the addition of
%the gluon with the lowest $k_{T}$ can never generate a net factor of $i\pi$
%since any such factors must cancel between the two diagrams where the lowest
%$k_{T}$ gluon lies either side of the cut.
 The first $i\pi$ terms and
the first  super-leading logarithm appear in case of four  soft gluons:%
\begin{equation}
\sigma_1\sim\sigma_{\text{Born}}\left(  \frac{2\alpha_{s}}{\pi}\right)  ^{4}\ln^{5}\left(
\frac{Q}{Q_{0}}\right)  \pi^{2}Y.\label{eq:sllform}%
\end{equation}
%Here, $Y$ is the size of the rapidity gap and $\sigma_0$ is the Born
%cross-section.
At higher orders in $\alpha_s$ more gluons can be outside the gap.
% At each order there is a maximum number of out-of-gap gluons that we
%expect to provide the maximum power of the logarithm stemming from the region where they all  become collinear to either of the initial state partons. 
%At each higher order we then have an additional
%$\alpha_s\ln^2(Q/Q_0)$ in such a configuration. What appears as  super-leading
%logarithms as compared to the single logs, $\alpha^n_s\ln(Q/O_0)^n$
%would thus constitute a series of  double logarithms,
%$\alpha^n_s(\ln^2(Q/O_0))^n$.
 However, to resum the double
logarithms to all orders a deeper understanding of the colour evolution of
multi-parton systems seems necessary.

Indeed we appear to have uncovered a breakdown of QCD coherence: radiation at
large angles does appear to be sensitive to radiation at low angles. 
%In fact, already  in case of the emission of a soft gluon from a pair of collinear partons 
However
this striking conclusion was arrived at under the assumption that it is
correct to order successive emissions in transverse momentum. Coherence
indicates that one does not need to take too much care over the ordering
variable, e.g. $k_{T}$, $E$ and $k_{T}^{2}/E$ are all equally good ordering
variables but the super-leading logarithms arise counter to the expectations
of coherence and in particular as a result of radiation which is both soft and
collinear. It is therefore required to prove the validity of $k_{T\text{ }}$
ordering before we can claim without doubt the emergence of super-leading
logarithms or confirm their size.
\section{Numerical results\label{sec:results}}

\begin{figure}[h]
\begin{center}
\centerline{\includegraphics[width=10cm]{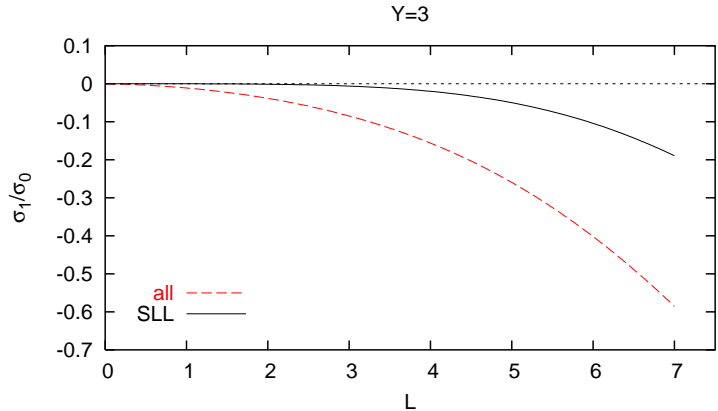}}
%\centerline{\includegraphics[width=10cm]{Y5SL.ps}}
\end{center}
\caption{$L$ dependence of the out-of-gap cross-section (normalized to the
in-gap cross-section) at Y=3\label{fig:Ydep}}%
\end{figure}

We numerically compute the out-of-gap cross section which is  the sum of Eq. (\ref{eq:real}) and Eq. (\ref{eq:virtual}) and which which we generically denote $\sigma_1$.  The purely superleading logarithmic part of $\sigma_1$ is obtained by considering the initial state collinear limit and performing the integral over rapidity over an interval of size $\ln(Q/k_T)$. The result is multiplied by 2 to account for the possibility that the out-of-gap gluon can be on either side of the gap. We refer to the cross section thus computed as `SLL' in the figures.  For comparison, we also compute the sum
of Eq. (\ref{eq:real}) and Eq. (\ref{eq:virtual}) without making the collinear
approximation. In this case the integral over $y$ is over the region
$Y/2<|y|<\Delta y/2+\ln(Q/k_{T})$ where $\Delta y = Y + 2$. This cross-sections is
labelled `all' and  necessarily includes a partial summation
of the single logarithmic terms as well as the super-leading terms. The strong coupling is fixed at $\alpha_{s}=0.15$.  Fig. \ref{fig:Ydep} shows $\sigma_1$ as a function of $L=\ln(Q^2/Q_0^2)$, normalized to the fully resummed cross-section $\sigma_0$  corresponding to zero gluons outside of the gap region, i.e.\ as determined by Eq. (\ref{eq:OS}).
% It seems that while the out-of-gap cross-section is not dominant anywhere it
%is also not negligible. This is of course already known:\ the non-global
%logarithms are generally significant. We can also see from these plots that
The super-leading series is generally small relative to the `all' result for
$L\lesssim4$, which indicates that the single logarithms are
phenomenologically much more important than the formally super-leading logs at
these values of $L$. Of course one should remember that our calculations are
for the emission of one gluon outside the gap region and the full
super-leading series requires the computation of any number of such gluons.

\begin{figure}[h]
\begin{center}
\centerline{\includegraphics[width=10cm]{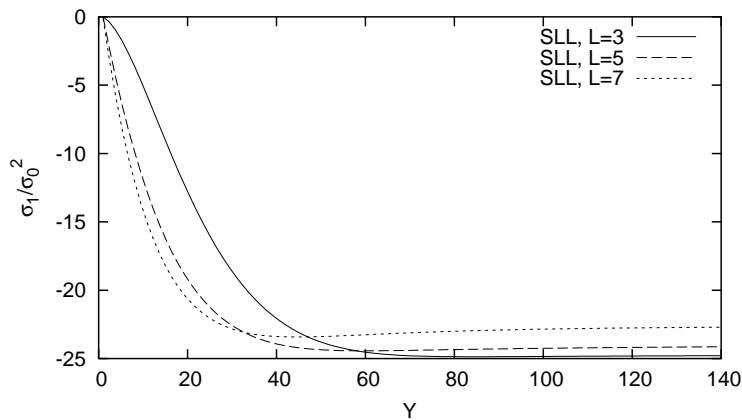}}
\end{center}
\caption{$Y$ dependence of the out-of-gap cross-section normalized to the
square of the in-gap cross-section at three different values of $L$ and
plotted out to very large $Y$.}%
\label{fig:largeYdep}%
\end{figure}

From a more theoretical perspective it is interesting to take a look at the
cross-sections out to larger values of $Y$, see Fig. \ref{fig:largeYdep}. Note that this time we have normalized the cross-section by the square of
the in-gap cross-section.  The cross-section saturates at large enough $Y$, i.e. $\sigma_{1}\sim-\sigma_{0}^{2}$. 
%This is particularly interesting with repect to the large-$Y$ behaviour of $\sigma_0$. 
In \cite{Forshaw:2005sx} we calculated the
conventional gap-between-jets cross-section in the high energy limit and
showed that it is equivalent to the BFKL result in the region in which both
are valid. Here, we find that in the high-energy (large $Y$) limit
the cross section for one emission outside the gap is proportional to the
square of the conventional gap cross-section, offering a tantalizing clue to
the structure of higher orders. A deeper understanding of this connection
would almost certainly open new avenues to understanding non-global observables.

\section{Outlook}
We appear to have found the breakdown of the intuitive picture of QCD coherence: superleading logarithms appear in the gaps-between-jets observable as the consequence of the sensitivity of soft wide angle gluon emission to collinear emission. The full confirmation of this finding though requires the proof of the validity of $k_T$-ordering.  The new super-leading contributions are not restricted to the gaps-between-jets observable. We expect them to arise generally
in non-global observables and potentially as additional leading logarithms also in global observables where non-global contributions are subleading. The new  contributions will  therefore  possibly have an impact on a wide spectrum of processes and observables, such as eventshape variables, $k_T$-distributions or particle production near threshold. The connection between the super-leading logarithms and high-energy QCD appears to offer intriguing clues for their resummation and the understanding of non-global observables.

%\item 
%One can view the gaps-between-jets process as a look at the pomeron
%loop in QCD, since one sums over radiation outside the gap
%(corresponding to a cut pomeron) and forbids it inside the gap
%(corresponding to one pomeron either side of the cut). 
%Studying the
%gaps-between-jets cross-section in the high energy limit (large $Y$)
%A deeper understanding of this connection could  provide valuable clues to a better understanding of 
%non-global observables. 
%\item 

%
%\section{Conclusion}
%
%\section*{Acknowledgements}
%I like to thank the organisers to enable this fruitful conference.
% and the organizers for a stimulating atmosphere.

\end{document}